\documentclass[aps,prl,amsmath,amssymb,reprint,floatfix]{revtex4-2}
\usepackage{lmodern}
\usepackage{graphicx}
\usepackage{dcolumn}
\usepackage{bm}
\usepackage[utf8]{inputenc}
\usepackage[T1]{fontenc}
\DeclareUnicodeCharacter{0394}{\ensuremath{\Delta}}
\DeclareUnicodeCharacter{03B1}{\ensuremath{\alpha}}
\DeclareUnicodeCharacter{03B2}{\ensuremath{\beta}}
\DeclareUnicodeCharacter{03BC}{\ensuremath{\mu}}
\DeclareUnicodeCharacter{03C3}{\ensuremath{\sigma}}
\DeclareUnicodeCharacter{03C0}{\ensuremath{\pi}}
\usepackage{mathptmx}
\usepackage{etoolbox}
\usepackage{xcolor}
\usepackage{booktabs}  
\usepackage{multirow}

\usepackage[english]{babel}

\usepackage{hyperref}
\begin{document}

\title{Orbital-weighted learning for transition metal complexes}
\author{Gabriel Klassen}%
\author{Guido Falk von Rudorff}%
\email{vonrudorff@uni-kassel.de}
\affiliation{Institut f\"ur Chemie, Universit\"at Kassel, Heinrich-Plett-Stra{\ss}e~40, 34132~Kassel,~Germany}
\affiliation{Center for Interdisciplinary Nanostructure Science and Technology (CINSaT), Heinrich-Plett-Stra{\ss}e 40, 34132 Kassel}

\date{\today}
\begin{abstract}
Frontier orbital energies are key to the description of molecular light-matter interaction and pose a challenge both for electronic structure calculations and for machine learning approaches, especially for transition metal complexes (TMC). Those are highly relevant since the gap between the highest occupied molecular orbital (HOMO) and the lowest unoccupied molecular orbital (LUMO) is in the range of visible light. For TMCs, gap position and magnitude are determined by metal center, its coordination and ligand identity. Since the systems exhibit a wide range of charge and spin states, frontier orbital energies can only be predicted with machine learning models that explicitly incorporate spin and charge. In this work, we show that a semi-empirical orbital population as weight rather than feature can increase efficiency for both physics-based and neural-network-based atomic representations. Because orbital populations are highly sensitive to both charge and spin, they offer a general pathway to render existing geometry-based machine learning representations useful for orbital energies. On TMC datasets, our approach reduces the required training data by up to 95\,\%, and improves physics-based representations to almost match the accuracy of pre-trained neural network representations at 12\,\% the inference compute cost. This greatly reduces the data and compute requirements to accurately model frontier orbitals in TMCs.
\end{abstract}

\maketitle

\section{Introduction}
The highest occupied molecular orbital (HOMO) and the lowest unoccupied molecular orbital (LUMO) together are the frontier orbitals. Both their absolute energy value as well as the difference, the so-called gap, are of fundamental interest to interactions between molecules and light, even as approximation\cite{Kimber2023,Kronik2012}. The HOMO energy is approximately the negative of the ionization potential, and as such describes approximately the amount of energy required to remove one electron from a molecule. The negative of the LUMO energy, again, is a proxy for the electron affinity, a measure of how easily a molecule would accept an electron. The orbital energies and the gap are relevant for applications such as OLEDs\cite{Aarti2025} or organic solar cells\cite{Yi2023}, but in theoretical modeling also finds use in reactivity analysis\cite{Miranda-Quintana2022,Pucci2022}.

Even though many quantum chemistry methods yield values for frontier orbital energies, the accuracy thereof varies widely, for example for transition metal complexes\cite{Dohm2018,Neugebauer2023}. With higher levels of theory being able to recover accurate values at higher cost, many researchers have investigated machine learning routes to enable accurate predictions at affordable cost, building upon density functional theory (DFT) data at GGA level in most cases\cite{Balcells2020}, as only few other data points exist\cite{Dohm2018,Garrison2026}. Unlike energies which can be approximated well as sum of local energy contributions, frontier orbitals are sensitive to the overall molecular structure, so they pose a harder problem even though their effective dimensionality is lower than the one of the energy\cite{Banjafar2025d}. Moreover,  applications regarding frontier orbitals often include charged molecules or a specific spin state for which many models have no native support\cite{Cho2026}.

One successful approach for dealing with highly diverse and clustered data sets is splitting data set into groups such as charge and spin\cite{Mazouin2022,Zeni2024,Lemm2023a,Srsen2024}, which is independent of the particular learning approach. Similarly, learning the difference between a computationally inexpensive and a more accurate higher level result, $\Delta$-ML\cite{Ramakrishnan2015}, and its generalization multi-fidelity learning has been applied to frontier orbitals and related properties such as excitation energies\cite{Lyu2025}.

Kernel or linear models require a representation or a descriptor of the system. Over the years, many such representations have been developed\cite{Langer2022,Musil2021}. Frontier orbitals or gaps have been learned using a wide range of geometry-based representations such as BAML\cite{Faber2018}, MTM\cite{Gubaev2018}, many-body expansions\cite{Pronobis2018}, MBTR\cite{Rahaman2020,Stuke2019}, cMBDF\cite{Khan2025a}, SPA$^\textrm{H}$M and its variants \cite{Briling2024,Fabrizio2022}, SOAP\cite{Kirschbaum2023}. For some application classes such as transition metal complexes (TMC), using the atomic representations of the metal center only\cite{Cho2026} proved to be an effective simplification. Some models extend existing representations such as SLATM with DFTB-derived properties\cite{HinostrozaCaldas2026} or built upon geometry free features from element embeddings\cite{Janet2017}, autocorrelation functions on the molecular graph\cite{Janet2017a,Nandy2018} or general graph features\cite{Zheng2023, Thinius2026,Hasan2025,Faber2017,Pinheiro2020,Orsi2025,Jorgensen2018,Pereira2017,Hung2025} including fragmentation approaches\cite{Mohamed2025}. Other approaches in similar spirit combine cheminformatics descriptors with electronic structure features through stacking\cite{Mahmoudi2026},  attention\cite{Zhang2026}, or concatenation\cite{Rahaman2020}. Fundamentally different is the idea to build representations from molecule-specific electronic structure calculations, as done in MOB-ML using Fock, Coulomb and Exchange matrix from a Hartree-Fock calculation which proved particularly helpful for post-SCF correlation energies\cite{Welborn2018,Cheng2019a,Husch2021} which has been similarly developed for general learning\cite{Karandashev2022}. This includes DFT-derived descriptors of metal centers\cite{Duan2019a} or the full system\cite{Caylak2021}, and learning a Löwdin-symmetrized effective Hamiltonian to reproduces MO energies\cite{Cignoni2024} and low-fidelity pre-training\cite{Kirschbaum2024}.

\begin{figure*}
    
    \centering
    \includegraphics[width=\linewidth]{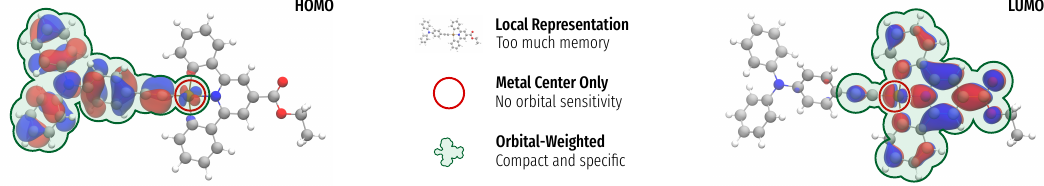}
    \caption{Illustration of the proposed method (orbital-weighted learning, OWL) for a transition metal complex. Traditional approaches either use all atomic environments (local) or the atomic environment of the metal center only (red circle). OWL uses weighted sums of the local environments of atoms with orbital population (green area)}
    \label{fig:owl}
\end{figure*}

For neural networks, several approaches have been considered: using semi-empirical Fock, density, Hamiltonian and overlap matrices as feature in OrbNet-Equi\cite{Qiao2022}, environments of top-$k$ atoms for multiple frontier orbitals in one representation (ESIN)\cite{He2025}, using a physics-inspired representation as input to a neural network\cite{Hou2018}, using a ZINDO-derived HOMO-weighted radial distribution function (MO-RDF) for TD-DFT in AttentiveFP\cite{Coxson2026}, learning the a weighted average pooling function for a MPNN with a second jointly-trained neural network\cite{Chen2023}, or transfer learning through augmentation of a feature with a low-fidelity label\cite{Buterez2024}. In a more general approach, learning the full wave function (e.g. in SchNOrb\cite{Schutt2019}) or orbitals (e.g. PaiNN\cite{Kirschbaum2023}) gives direct access to the frontier orbital energies and allows further analysis of the effective many-body-character\cite{Esders2025}. 

Other approaches not including electronic structure-based features are described in comprehensive reviews\cite{Nandy2021,MantillaSantaCruz2026}.

\section{Methods}
Many molecular representations exist in two variants: global and local. In the global version, the information of a molecule is collected in a single feature vector, while the local version contains a feature vector for every atom. The atomic feature vectors then form a matrix with the shape $n_\text{features}\times n_\text{atoms}$ for every molecule. While a local representation usually yields better performance for Kernel-Ridge-Regression (KRR) models for geometry-based properties, it comes with higher computational cost and hardware requirements, such as increased memory usage which may require batching\cite{Grimblat2026}.

Since frontier orbitals in TMCs are typically localized around a metal center, their character depends on the surrounding ligands, making them quite challenging to handle. As most of the physical representations scale with the number of unique elements in a dataset\cite{Darby2022,Musil2021} and the average number of atoms in TMCs is high (see Tab.~\ref{tab:datasets}), the usage of local representations becomes impractical and for some long representations  infeasible: some representations such as SLATM reach over a million features on a typical TMC dataset\cite{Cho2026}. To build a model that can predict the energy of frontier orbitals in large, chemically diverse systems, it is therefore advantageous to use global representations, which can be seen as a pooling of local representations\cite{Chen2023,De2016}. For frontier orbitals, we propose a weighted sum of local representations where the weights are orbital coefficients from a semiempirical calculation. After the weighted sum, we employ a standard KRR model across different datasets and compare them with other state-of-the-art models such as pre-trained MACE variants, see the schematic in Fig.~\ref{fig:owl}.

Unlike augmenting existing geometric representations with orbital features, the weighting offers a more direct way of grouping training points in feature space for this application, since small variations in geometry can yield a large discontinous response in orbital shape, e.g. through crossing of orbital energies. Using geometry information alone would induce an errorneous similarity.

\subsection{Orbital Weighting}
When sum-pooling a local representation into a global version, the features of all atoms would contribute equally to the global vector. Since the frontier orbital is localized for most TMCs, the na\"{i}ve sum will dilute the signal of the relevant atomic environments, leading to an effect similar to noise in the representation. To identify the atoms that carry a higher contribution to the formation of the frontier orbitals, we perform a Mulliken population analysis of one specific orbital only. By selecting a specific molecular orbital (MO) $i$, we can obtain an expression for the weight $w_A$ of atom $A$ to this MO as
\begin{equation}
    w_A^{i} = \sum_{\nu \in A}\sum_\mu \mathbf{C}_{\mu i}\mathbf{S}_{\mu \nu}\mathbf{C}_{\nu i}.
    \label{eqn:Weights}
\end{equation}
with standard coefficient matrix $\mathbf{C}$ and overlap matrix $\mathbf{S}$. The sum over $\nu \in A$ runs over all atomic orbital (AO) functions centered on atom $A$, while $\mu$ runs over all AO in the molecule. With those weights, we obtain a global representation $\mathbf{r}^i$ for each MO $i$ from all generic atomic representations $\mathbf{r}_A$:
\begin{equation}
    \mathbf{r}^i=\sum_\text{A}w_A^i\mathbf{r}_\text{A}
    \label{eqn:Global_rep}
\end{equation}
This can be seen as a general expression to obtain a global representation from a local version. For sum-pooling, all weights are one, with every atom weighted uniformly. Using the metal center only would mean to have all weights zeros, except for one entry corresponding to the metal which would be of unit weight. 
To obtain the coefficients and the overlap matrix in Eq.~\ref{eqn:Weights}, we calculate the orbital energies with the semi-empirical method GFN2-xTB and, in the case of open-shell systems, the spin-polarized version thereof \cite{Neugebauer2023}. While higher levels of theory would yield a different weight vector that might be more accurate, we chose xTB to balance computational cost with physically reasonable results. Nothing in our approach is restricted to this particular level of theory.

\subsection{Charge and spin detrending}
Detrending and centering data is a common practice in machine learning to remove a systematic offset. When a dataset contains molecules with different charges, the frontier-orbital distribution shifts and becomes multimodal, since molecular orbitals are sensitive to charge. The multimodal distribution tends to have higher variance, and representations without a charge and spin embedding will not capture this offset effectively. The effect of the spin might not be as strong as the charges, but for TMCs both properties are often coupled, and both will have an effect on the frontier orbitals. To detrend these properties, we collect the charges and spin states for every molecule in the dataset via one-hot encoding in a design matrix $\mathbf{A}$. With this matrix, we then perform a least-squares fit on the training data
\begin{equation}
    \mathbf{x}^\ast = \arg\min_{\mathbf{x}}
    \lVert \mathbf{A}_{\text{train}}\mathbf{x} - y_{\text{train}} \rVert_2^2
\end{equation}
This fitted contribution $\mathbf{Ax}^\ast$ is then subtracted to obtain the detrended residuals $\tilde{y}$
\begin{equation}
    \tilde{y} = y - \mathbf{A}\mathbf{x}^\ast
\end{equation}
We then center the residuals around zero by subtracting the overall mean. The residuals then show almost no charge- and spin-dependent effects and reduced variance.

\subsection{Kernel ridge regression}
To test our method, we employed Kernel ridge regression (KRR), where a kernel function $k(\mathbf{x},\mathbf{x}')$ measures the similarity of data points. We use the Gaussian kernel. In this work, this refers to the similarity between representations $\mathbf{x}$. To predict the label of a query point $\mathbf{x}_\textrm{q}$
we use the form
\begin{align}
    f(\mathbf{x}_\textrm{q}) = \sum_{i=1}^{n_t} \boldsymbol{\alpha}_i k(\mathbf{x}_\textrm{q}, \mathbf{x}_i)\label{eqn:globalinterpolant}
\end{align}
where the model coefficients $\boldsymbol{\alpha}$ are obtained via direct inversion 
\begin{align}
    \boldsymbol{\alpha} = (\mathbf{K}+\lambda \mathbf{I})^{-1}\mathbf{y}
    \label{eqn:krr}
\end{align}
with kernel matrix element $\mathbf{K}_{ij}=k(\mathbf{x}_i, \mathbf{x}_j)$ and  regularization strength $\lambda$. Hyperparameter optimization and cross-validation follow previous work\cite{Grimblat2026}.

\subsection{Datasets and Representations}
To test our modifications on representations, we evaluate them on FCHL19\cite{Christensen2020a}, cMBDF\cite{Khan2025a}, and MACE\cite{Batatia, Shiota2024}. MACE is built as a neural network potential, but a pre-trained version of the model can also be used as a (local) representation\cite{Shiota2024}. In this work, we use pre-trained versions of MACE: MACE-OFF, trained on small organic molecules\cite{Kovacs2025}; MACE-MP0, trained on the MPtrj dataset\cite{Batatia2024}; and MACE-OMol, trained on the OMol25 dataset\cite{Levine2025}. For MACE-OFF and MACE-MP0, we use the \textit{medium} subvariant. The MACE-OMol version also features charge and spin embedding.

We tested on the tmPHOTO, TM-GSspin$^+$\cite{Cho2026}, and QM9 datasets. The tmPHOTO dataset is a subset of the tmQM\cite{Kevlishvili2025} database and consists of photoactivity-associated complexes with a charge from $-1$ to $+1$ in a singlet spin state. The TM-GSspin$^+$ dataset consists of TMCs with charges from $-5$ to $+4$ and spin multiplicity up to 6. The QM9 dataset is a well-established benchmark set for ML models and consists of organic molecules with up to nine heavy atoms. Further details of the dataset are shown in Tab.~\ref{tab:datasets}.

\begin{table}[h]

\caption{Descriptive statistics of the datasets used in this work.}
\begin{ruledtabular}
\begin{tabular}{llll}
\label{tab:datasets}
Dataset           & tmPHOTO & TM-GSspin & QM9      \\\hline
Training points    & 4,268    & 2,260      & 133,885  \\
Unique elements   & 25      & 18        & 5        \\
Charges           & -1 to 1 & -5 to 4   & none     \\
Spin multiplicity & 1 to 6  & singlet  & singlet \\
Average atom count & 73 & 54 & 18\\
\end{tabular}
\end{ruledtabular}
\end{table}

\section{Results}

\subsection{Photoactive transition metal complexes}
Frontier orbital energies are important for the photophysical properties of complexes, such as absorption and emission. A suitable dataset is the tmPHOTO dataset, a subset of tmQM. This subset was derived from the original dataset by filtering for complexes that appeared in manuscripts related to photophysical activity\cite{Kevlishvili2025}. When testing our method on the tmPHOTO dataset, we applied the same element filtering as in literature\cite{Cho2026}. The remaining dataset contains 4,268 complexes with 25 unique elements. 

\begin{figure*}
    \centering
    \includegraphics[width=1\linewidth]{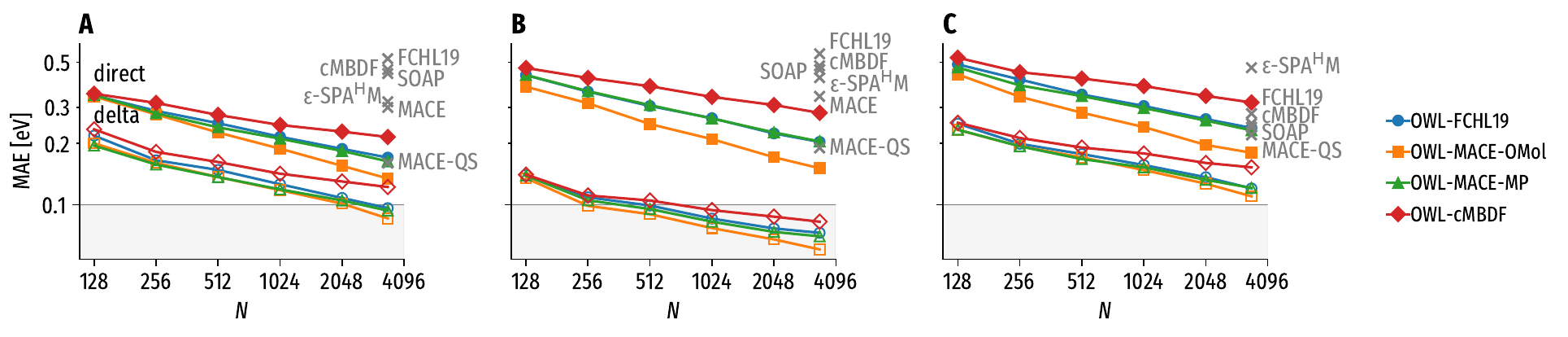}
    \caption{Learning curves for the tmPHOTO dataset\cite{Kevlishvili2025} after inclusion of the proposed orbital weightings, both for direct (filled markers) and $\Delta$-learning (empty markers) with an xTB baseline. Data shown for HOMO (A), LUMO (B) and gap (C) up to $80\%$ of the dataset (3,414 training points). The target accuracy of $0.1\, \mathrm{eV}$ is shaded in grey. The reference values (grey crosses) show models\cite{Cho2026} without the proposed method.}
    \label{fig:Learningcurve_tmPHOTO}
\end{figure*}

With our orbital-weighted learning (OWL), every property (HOMO, LUMO and gap) is learned more efficiently, see Fig.~\ref{fig:Learningcurve_tmPHOTO}. For HOMO, using orbital-weighting makes both physics-based representations FCHL19 and cMBDF more accurate with 95\,\% fewer data (128 data points instead of 3,414, i.e. 80\,\% of the dataset). Large pre-trained models are only slightly better than physics-based representations for HOMO energies in both the direct and $\Delta$ learning case. While this indicates that the pre-training on extensive datasets does not yield a material improvement, using KRR as additional head/readout function during MACE training may further improve the models. The improvement for frontier orbitals is a consequence of OWL being highly sensitive to the molecular charges. Since the dataset contains three different molecular charges, the raw frontier-orbital distribution is multi-modal, which constitutes a harder learning target. This applies particularly to kernel ridge regression which reverts to predicting the mean label for query points far away from any training data. This is in line with earlier observations\cite{Cho2026} that using MACE-QS (a MACE variant which includes charge and spin embeddings) performs significantly better than MACE alone.

For the LUMO in Fig.~\ref{fig:Learningcurve_tmPHOTO}B, we observe the same pronounced improvement in data efficiency and OWL-FCHL19 and OWL-MACE-MP are close in performance. Only OWL-MACE-OMol shows higher data efficiency requiring about 25\,\% of the training data that other representations need. All representations for all properties benefit from our orbital weighting. As one would expect from learning the difference of two separate functions, the learning curves for the gap in Fig.~\ref{fig:Learningcurve_tmPHOTO}C follow the behavior of the quantity that is harder to learn in the first place, the LUMO energy. Since we have to choose either the HOMO or LUMO weights when learning the gap, neither of these options on its own is solely representative of the gap character. We find that for direct learning LUMO yields better results and for $\Delta$ learning, the HOMO weights perform better. Remarkably, the prediction accuracy improves by using the weights even for predicting the gap.

Since our method requires an xTB single-point calculation, a natural extension to direct learning is a $\Delta$-learning approach between the target's higher level of theory and the xTB orbital energies. For the HOMO energy, we obtained an MAE similar to MACE-QS, the best-performing model on this dataset, when applying our method to the physics-inspired representation FCHL19. We achieved the overall lowest MAE with the data-driven representation MACE-OMol in our KRR setting. For all data-driven pre-trained representations, the effective training data available to the pre-trained models is much larger: the OMol dataset contains $140$ million data points\cite{Levine2025}. Consequently, their use constitutes transfer learning. For physics-inspired representation FCHL19, all improvement is achieved without any pre-training. 

In the $\Delta$-learning approach, three of the four representations were able to achieve an MAE below $0.1\,\mathrm{eV}$ between 2,048 and 3,414 datapoints. Only cMBDF did gain little across all tests, since it is built to exhibit a constant length, independent of the number of unique elements in the dataset and the element identity only enters through the elemental kernel function. Our summation therefore suffers from information loss that impairs its learning behavior. This is evident in the flattening of the learning curves for larger training set sizes. The learning curves for the LUMO energy in direct learning are comparable to the curves for the HOMO. With the $\Delta$ approach however, it shows a significantly lower MAE, with all the representations reaching the $0.1\,\mathrm{eV}$ threshold with less than 1,024 training points. To investigate the different behavior of the learning curves, we examined the correlation between the xTB energies and the target values at the TPSSh-D3(BJ)/def2-SVP level and observed a higher correlation for the LUMO than for the HOMO energy. Due to this difference in correlation, the LUMO-$\Delta$ is the easier-to-learn property compared to the HOMO-$\Delta$. 

The multi-modal character of the frontier orbital energy as consequence of the charge state is greatly reduced for the HOMO-LUMO gap, leading to better relative performance of the representations without charge embedding in Fig.~\ref{fig:Learningcurve_tmPHOTO}C. While cMBDF yields a higher MAE with our method due to information loss in the summation, OWL improves the performance of FCHL19 by reducing the data requirements by 75\,\%. When using the weights corresponding to the LUMO, FCHL19 shows a 15\,\% lower MAE, while the usage of the weights corresponding to the HOMO yields a 5\,\% lower MAE. For every tested representation, using the LUMO weights in direct learning yields the lower MAE. Interestingly, for the $\Delta$-learning approach, this pattern reverses; here, every representation benefits from using the HOMO weights. While a strong offset effect from using $\Delta$ learning with only flat learning afterwards is documented for the strict superset tmQM\cite{Kneiding2026}, we observe continuous learning for tmPHOTO even after reducing the offset.
When both weights are applied subsequently to the representation, the models perform worse than with a single weight representative of either HOMO or LUMO, so we always choose either MO. In Fig.~\ref{fig:Learningcurve_tmPHOTO}C, we show the best case for each curve. For MACE-MP0, we measured an improvement similar to FCHL19.

\subsection{TMCs with higher spin multiplicity}
Another important feature of TMCs is that they occur in various spin multiplicities. Many established representations have been developed for and benchmarked on small organic molecules and, hence, do not embed spin information. In OWL, the weighting naturally depends on the electronic state including spin and charge.

To test our method across different spin multiplicities, we chose the TM-GSspin$^+$ dataset, an extension of the TM-GSspin dataset designed to build models related to spin-related properties \cite{Cho2024, Cho2026}. 
TM-GSspin$^+$ yields results similar to those of tmPHOTO. While the MAE is higher for TM-GSspin$^+$, the improvement due to OWL is comparable. For both TMC datasets, the MAE of the frontier orbital prediction could be reduced significantly, except for MACE-OMol, where the performance improvement is only 15\,\% for the HOMO and 21\,\% for the LUMO energy. The TM-GSspin$^+$ dataset shows greater diversity in spin multiplicity than the tmPHOTO dataset, but by using spin-polarized xTB\cite{Neugebauer2023}, our method shows a consistent improvement with every representation we tested. The relative improvement on the spin dataset is slightly higher than on tmPHOTO, with MAE reductions up to 70\,\% for FCHL, 74\,\% for MACE-MP, and 59\,\% for cMBDf. MACE-OMol remained rather constant. The higher MAE shown in Fig.~\ref{fig:Efficiency-plot} for TM-GSspin$^+$ compared to the other datasets is a combination of the higher complexity due to spin multiplicities and the smaller training set size: our largest training set size of 80\,\% of the dataset corresponds to $3,414$ training points for tmPHOTO and $1,808$ training points for GSspin$^+$.

\subsection{Small organic molecules}
Most physics-based representations were developed on datasets consisting of relatively small molecules with only a handful of unique elements. These representations perform well e.g. on QM9 and similar datasets. We do not expect OWL to be particularly effective for small systems, since the frontier orbitals are rarely localized only on small parts of these systems. To explore the limits of OWL, we tested the prediction accuracy on QM9. In this case, we replaced MACE-MP0 with MACE-OFF, as this variant was pre-trained on organic molecules similar to QM9 and performs better than the MP0 variant on such datasets. In our tests, most representations performed worse on this dataset when OWL is included. MACE-OMol again had the lowest MAE of all representations, and the effect of our method was minor but still slightly negative, with a 1\,\% improvement in the HOMO, a 7\,\% decrease for the LUMO, and a 16\,\% decrease for the gap. For cMBDF, the MAE for the HOMO increased by 23\,\% with our method; the gap and LUMO predictions showed a slight decrease in MAE of 3\,\% and 8\,\%, respectively. For FCHL19, we saw a noticeable improvement in the LUMO (33\,\%), a moderate improvement in the gap (19\,\%), and a minor improvement in the HOMO (8\,\%). The biggest outlier is MACE-OFF; here we observed a significant increase in the MAE for every property when applying our method. The MAE for the HOMO and gap predictions doubled, while the MAE for the LUMO increased by 47\,\%. This aligns with earlier work showing that scaling and compressing features for MACE as a representation is ineffective\cite{Banjafar2026a} on QM9. However, the decrease in efficiency for our method stems from the dataset structure itself. The TMC datasets contain large molecules with an average size of 73 atoms per molecule for tmPHOTO and 54 atoms per molecule for TM-GSspin$^+$, and a single metal center. Therefore, the frontier orbitals are relatively localized in TMCs. Our method exploits this localization; QM9, by contrast, contains smaller molecules with 18 atoms per molecule. Because of the size and structural differences between the organic molecules and the TMCs, frontier-orbital localization is less pronounced, which decreases the effectiveness of our method on datasets such as QM9.

\subsection{Overall effects}
\begin{figure}[h]
    \centering
    \includegraphics[width=1\linewidth]{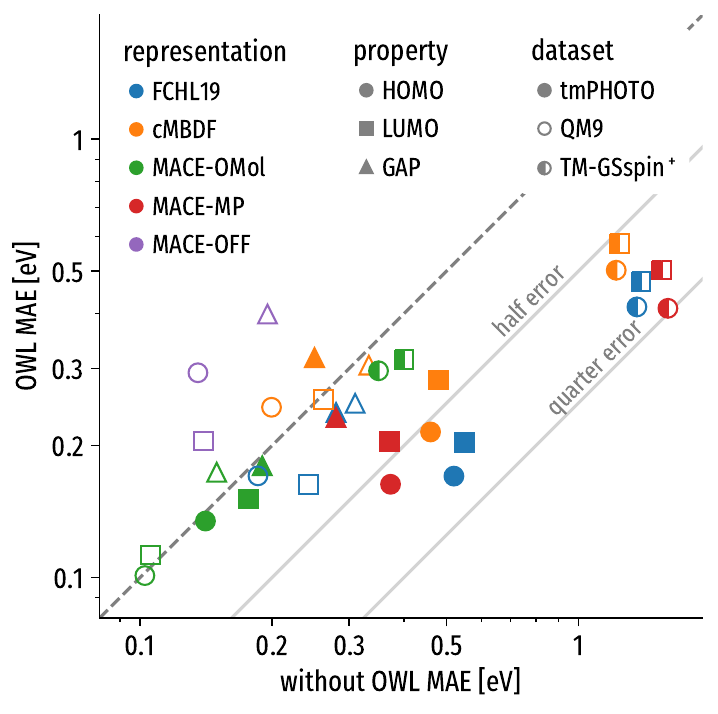}
    \caption{MAE with and without orbital-weighted learning (OWL) for all direct learning cases of the representations (colors), datasets (filling) and properties (shape). The data is shown for $80\%$ of the dataset for tmPHOTO and TM-GSSpin$^+$ and $4,096$ training points for QM9. Points below the diagonal show an improvement, while points above show a decline in performance.}
    \label{fig:Efficiency-plot}
\end{figure}

For a systematic comparison, we collect models for all representations on the same data with and without using our method OWL, see Fig.~\ref{fig:Efficiency-plot}. In general, we observe OWL to improve data efficiency for large molecules and orbital energies. Interestingly, MACE-MP benefits substantially from OWL, more than halving the error. Since that model is pre-trained on materials and not molecules, OWL supports its transfer onto molecular systems without additional training, thus increasing its applicability domain. This is likely due to the strong influence of the metal-center on the orbital character, where the metal center is closer to the training data of MACE-MP.

For physics-inspired representations, OWL typically halves the error, in some cases reduces it to a quarter as shown in Fig.~\ref{fig:Efficiency-plot}. This can be understood since the many-body interactions for large systems involve atoms which do not influence the orbitals strongly. This behaves like an invariance: along changes of molecular structure or atomic environment far away from the orbital in question, neither orbital shape nor energy are subject to substantive change. This invariance then needs to be discovered by the models from the data alone, which makes those models less data efficient. If those of the many-body contributions which are less relevant are suppressed by the weighting, this inefficiency is removed.

As shown in Fig.~\ref{fig:Learningcurve_tmPHOTO}, the learning curves for HOMO, LUMO and gap are all very similar, although shifted. That is the because the label variance after categorical detrending by charge state in tmPHOTO is ordered with HOMO$<$LUMO$<$gap. The data efficiency (i.e. the learning curve slope) is comparable for all models, so only the offset originating from the label variance is different. This trend is reproduced for OWL in Fig.~\ref{fig:Efficiency-plot}, where HOMO and LUMO benefit either both or none from the orbital weighting, indicating that the relevance selection induced by this method yields consistent results for relevant orbitals.

The weights in OWL are approximate, since they are obtained at low computational cost at semi-empirical level only. This assumes a strong similarity of the orbital characters in both methods which is validated by the success of the method only. In Fig.~\ref{fig:Efficiency-plot}, this seems to be the case even for transition metal complexes where semi-empirical methods cannot offer a reliable quantitative assessment of different spin states. It is to be expected that the support from the semi-empirical weights then helps inaccurate models to improve but offers little improvement in the limit of already accurate models. This is observed in Fig.~\ref{fig:Efficiency-plot}, where OWL seems to improve the models to about 0.15\,eV. In the limit of infinite training data, the benefit from OWL vanishes the same way as the one of $\Delta$-learning does.

\subsection{Timings}
\begin{figure}[h]
    \centering
    \includegraphics[width=1\linewidth]{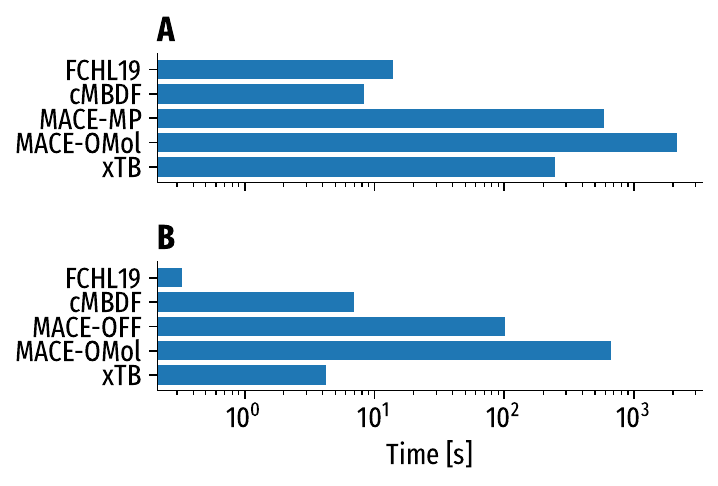}
    \caption{Time required to calculate the different representations for 500 molecules on tmPHOTO (A) and QM9 (B) on a single core. } 
    \label{fig:timings}
\end{figure}

The time required to use our method is the sum of the time required to first calculate the geometry-based representation and then the time to perform a semi-empirical calculation. The time to apply the weight vector to the representation is negligible. Fig.~\ref{fig:timings} shows the timings for a random subset of 500 molecules each from the tmPHOTO and QM9 datasets. The xTB calculations were performed with the \texttt{tblite} Python package (version 0.6.0) using spin-polarized GFN-2 xTB. We performed the time measurements on a single CPU core to enable a fair comparison of the total compute effort. The neural networks would benefit from GPU acceleration while FCHL and cMBDF would benefit from multicore CPU parallelization.

The data-driven representations require significantly more time to calculate, with MACE-OMol taking the most time due to its parameter count. While this representation on average performed better than the geometry-based representations for direct learning tasks, its performance for $\Delta$-learning is comparable to other representations, as shown in Fig.~\ref{fig:Learningcurve_tmPHOTO}. On the tmPHOTO dataset, 500 molecules required $2,163$ seconds for MACE-OMol. FCHL19 takes $14$ seconds instead, and the xTB single-point calculations took $406$ additional seconds. The compute cost does not increase for multiple frontier orbitals, since all weights for all MOs are obtained from the same calculation. In our KRR workflow, inference time is dominated by calculating the representations rather than by model inference\cite{Grimblat2026}. Using orbital weighting benefits from the fast calculation time of geometry-based representations while keeping the extra computational time to a minimum by using xTB for the orbital weights. 

\section{Conclusions}
In this work, we demonstrated that the applicability of geometry-based representations can be extended to large and diverse transition metal complexes (TMCs) by including orbital information. With only geometry information, spin and charge of the TMCs cannot be resolved by the representation. Conceptually, ligand parts which do not contribute to the orbital in question form an invariance in feature space, since many modifications of that part of the molecule would not change the orbital shape or energy significantly. With our orbital weighting, we scope the feature space to only the relevant parts of the molecule. Interestingly, both physics-based and neural-network representations benefit. 

With orbital weighting learning (OWL), TMCs of different spin and charge states can be trained jointly even if the original geometry-based representation is not adapted for charge and spin. Orbital weighting performs best if large molecules with partially localized frontier orbitals are considered, such as TMCs. The improvement over existing approaches is most prominent for cases with large element diversity and applications with many charge and spin combinations. For small molecules where frontier orbitals are distributed over essentially the full molecule, OWL offers no advantage.

OWL inference requires an approximate molecular orbital expressed in per-atom coefficients. Although we use semi-empirical methods in this work, this is not a requirement, as orbital populations are commonly available. Geometry-based representations with OWL are much faster than neural-network models but regularly achieve comparable performance. This reduces inference time, but leaves the semi-empirical method as the current bottleneck.

\bibliography{nablachem} 
\end{document}